\documentclass[sigconf]{acmart}
\setcopyright{none}
\renewcommand\footnotetextcopyrightpermission[1]{}
\usepackage{booktabs}
\usepackage{tabularx}
\usepackage{microtype}
\usepackage{xcolor}
\usepackage{hyperref}
\usepackage{enumitem}
\usepackage{tikz}
\usetikzlibrary{arrows.meta,positioning,calc,shapes.geometric,fit,backgrounds}
\newcommand{\schele}[1]{\textsf{#1}}

\title{Inference-Aware \& Privacy-Preserving Deletion in Databases}

\author{Vishal Chakraborty}
\affiliation{\institution{University of California, Irvine}
\country{USA}
}
\email{vchakrab@uci.edu}

\author{Youri Kaminsky}
\affiliation{\institution{Hasso Plattner Institute,\\ University of Potsdam}
\country{Germany}
}
\email{youri.kaminsky@hpi.de}

\author{Arnav Abhijit Dhariya}
\affiliation{\institution{University of California, Irvine}
\country{USA}
}
\email{dhariyaa@uci.edu}

\author{Sharad Mehrotra}
\affiliation{\institution{University of California, Irvine}
\country{USA}
}
\email{sharad@ics.uci.edu}

\author{Felix Naumann}
\affiliation{\institution{Hasso Plattner Institute,\\ University of Potsdam}
\country{Germany}
}
\email{felix.naumann@hpi.de}

\author{Sarvesh Pandey}
\authornote{Work done while the author was a Fulbright Visiting Scholar at University of California, Irvine.}
\affiliation{\institution{Banaras Hindu University, Varanasi}
\country{India}
}
\email{sarveshpandey@bhu.ac.in}

\renewcommand{\shortauthors}{Chakraborty et al.}

\setlist[itemize]{leftmargin=1.2em,topsep=2pt,itemsep=1pt,parsep=0pt}
\setlist[enumerate]{leftmargin=1.4em,topsep=2pt,itemsep=1pt,parsep=0pt}

\begin{document}

\begin{abstract}
Deletion is a fundamental database operation, yet modern systems often fail to provide the privacy guarantee that users expect from it. 
A deleted value may disappear from query results and even from physical storage while still being inferable from dependencies, derived data, or traces exposed by the deletion event itself. 
Meaningful deletion, therefore, requires more than logical removal or physical erasure; it requires a privacy guarantee that limits what remains inferable after deletion. 
In this paper, we take an inference-centric view of deletion, focusing on two leakage channels: leakage from the post-deletion state and leakage from the deletion pattern itself. 
We use this lens to distinguish logical, physical, and semantic deletion, organize the design space of deletion operations, and highlight open research challenges for building deletion mechanisms with meaningful privacy guarantees in database systems.
\end{abstract}

\maketitle

\section{Introduction}

Deletion in databases has traditionally been treated as a well-scoped action: removing a record or value from logical state, and eventually from physical storage. In modern data systems, however, deletion is no longer a single operation, but a family of transformations over many artifacts: base tables, views, caches, indices, logs, summaries, and learned models. A privacy-driven request such as ``delete my salary’’ must therefore contend not only with the target cell, but also with dependencies that make the salary re-inferable from retained attributes, and with derived artifacts~(views, learned AI models, etc.) that may already encode its information~\cite{BuildingDelCompDB, chakraborty2023data, Chakraborty2024}. The result is a gap between deletion as removal and deletion as a privacy guarantee.

This gap matters more than ever. Data regulations have elevated the concept of deleting personal data from a storage-management feature to a compliance obligation~\cite{Chakraborty2024datacase}. Yet removing a value from query results, or even reclaiming its bytes from storage, does not erase the information that value contributed. In practice, deleted information may persist through what the system still exposes. In our view, this residual exposure arises through two broad channels: the remaining visible state and the deletion footprint itself. The result is a mismatch between what users, regulators, and applications expect from deletion and what database systems actually deliver.

\begin{example}
\label{ex:employee}
A company maintains employee records~(Figure~\ref{tab:employee-example}) with attributes
\schele{EID}, \schele{Team}, \schele{Title}, \schele{Level},
\schele{Manager},\schele{Location}, and \schele{SalaryBand}. Dependencies D1-D3 express which data are dependent. Suppose we want to delete the \emph{target cell}
$c^*=e_5[\schele{SalaryBand}]$.
\end{example}

\begin{figure*}[t]
\centering

\renewcommand{\arraystretch}{.8}

\begin{minipage}[t]{0.45\textwidth}
\centering
\setlength{\tabcolsep}{1pt}
\begin{tabular}{@{}lccccccc@{}}
\toprule
\schele{EID} & \schele{Team} & \schele{Title} & \schele{Level} & \schele{Manager} & \schele{Location} & \schele{SalaryBand} \\
\midrule
$e_1$      & Ads   & SWE   & L4 & $m_{10}$ & NYC & B4 \\
$e_2$      & Ads   & SWE   & L4 & $m_{10}$ & NYC & B4 \\
$e_3$      & Ads   & SWE   & L5 & $m_{10}$ & NYC & B5 \\
$e_4$      & Infra & SRE   & L4 & $m_{12}$ & SEA & B4 \\
$e_5$ & Ads & SWE & L4 & $m_{10}$ & NYC & \textcolor{red}{\textit{\textbf{B4}}} \\
\bottomrule
\end{tabular}
\end{minipage}
\hfill
\begin{minipage}[t]{0.54\textwidth}
\centering
\footnotesize
\setlength{\tabcolsep}{3pt}
\begin{tabular}{@{}l c c@{}}
\toprule
\textbf{ID} & \textbf{Dependency} & \textbf{Intuition} \\
\midrule
D1 & $(\schele{Team},\schele{Title}) \Rightarrow \schele{SalaryBand}$ &
Pay bands are standardized by job family within a team \\
D2 & $\schele{Level} \Rightarrow \schele{SalaryBand}$ &
Leveling ladders map levels to narrow compensation bands \\
D3 & $\schele{Manager} \Rightarrow \schele{Level}$ &
Managers supervise within limited level ranges \\
\bottomrule
\end{tabular}
\end{minipage}

\caption{Database instance and dependencies used in Example~\ref{ex:employee}.}
\label{tab:employee-example}
\end{figure*}

To delete the target cell $c^*$~( Example~\ref{ex:employee}), in a conventional system like PostgreSQL~\cite{PostgreSQL} one would remove tuple $e_5$ using \texttt{DELETE} and then insert the tuple $(e_5, \schele{Ads},\schele{SWE}, \schele{L4}, \schele{m}_{10}, \schele{NYC}, \schele{Null}).$ This removes $c^*$ from the logical state. 
But, this by itself, does not resolve the semantic question: \emph{after the
target value is removed from the visible state, what information about it
remains inferable from what the system retains and exposes?}

Consider Example~\ref{ex:employee} again. The dependencies in
Figure~\ref{tab:employee-example} creates multiple inference paths that can be used for inferring deleted values. When \schele{SalaryBand} is set to \texttt{Null}, it may be inferred directly through dependency D1 or D2, and transitively
through D3$\circ$D2:
$\schele{Manager}\Rightarrow\schele{Level}\Rightarrow\schele{SalaryBand}$.
Logical deletion does not ensure that its information content has
disappeared from the system.

Physical deletion does not solve the problem either: deleted tuples may linger in MVCC~\cite{postgres-mvcc-intro-doc} systems until reclamation runs (e.g., until \texttt{VACUUM} executes in PostgreSQL~\cite{postgres-routine-vacuuming-doc}), or persist as tombstones in LSM-based engines until compaction~\cite{Lethe}. Timely byte reclamation is itself an open problem~\cite{BuildingDelCompDB,Sarkar2022}, though one orthogonal to our focus. Even perfect reclamation does not ensure the deleted value is no longer inferable from what remains.

One may then ask if the answer is to delete more dependent data.
However, such attempts, which involve deleting \emph{all} data belonging to a user, as explored in systems such as DELF~and K9db~\cite{DakAlbab2023,DelFB}, are not
satisfactory by themselves: they can lead to a large amount of additional deletion, trigger
downstream maintenance, while remaining unclear what semantic guarantee is
being provided. 

A more selective alternative, perhaps, is to delete additional data
specifically to block inference paths created by dependencies. While this addresses an important part of
the problem, it opens further challenges. First, blocking all
re-inference paths may require many auxiliary deletions beyond the target
cell, degrading data utility. Second, the \emph{pattern} of those auxiliary
deletions can itself become a side channel. Consider Example~\ref{ex:employee} again.  
Suppose that dependency $D1$ is \emph{value-dependent} and predictive only for salary bands B3 and above. If the mechanism deletes \schele{Team} and \schele{Title} to block that path, the deletion footprint itself reveals that the hidden value likely belongs to the range where this dependency is active. Thus, auxiliary deletions may reveal which value the target cell may have had.

These observations suggest that a deletion mechanism should be evaluated
along at least three dimensions: a meaningful \emph{semantic guarantee}
about what remains inferable and through which channels; acceptable
\emph{deletion cost}, since auxiliary deletions reduce utility and alter
downstream artifacts; and acceptable \emph{systems overhead}, since
deletion interacts with maintenance, refresh, reclamation, compaction, and
other operational processes that determine when and how erasure takes
effect. In our view, these requirements are best understood by treating deletion as
an \emph{inference-control problem}.

Our work has taken a first step in this direction and explored how to formalize deletion relative to a declared dependency model~\cite{Chakraborty2024}. 
\emph{Our work shows that meaningful deletion not only admits a formal interpretation beyond logical removal, but also underscores that the broader problem space remains fragmented.}


\begin{figure}[b]
\centering
\begin{tikzpicture}[x=1cm,y=1cm, font=\small]
  \begin{scope}[xscale=0.8]

  \draw[->, line width=0.7pt] (0,0) -- (8.5,0) node[ right] {time};
  \draw[->, line width=0.6pt] (0,0) -- (0,3.55);
  \node[rotate=90, anchor=south] at (-0.15,1.775) {\textcolor{blue!70}{inference about $c^*$}};

  \fill[blue!6] (0,0.55) rectangle (7.55,0.85);
  \node[anchor=west,gray!70!black] at (0.12,0.70) {\scriptsize baseline prior};

  \def\xins{0.85}
  \def\xlog{2.35}
  \def\xphy{4.00}
  \def\xsem{6.10}
  \def\xp2e2{6.75}
  \def\xdelall{8}

  \def\yins{1.30}

  \def\ylog{2.85}

  \def\yphy{2.40}

   \fill[black] (\xins,0) circle (1.5pt);
   \fill[black] (\xlog,0) circle (1.5pt);
   \fill[black] (\xphy,0) circle (1.5pt);
   \fill[black] (\xp2e2,0) circle (1.5pt);
   \fill[black] (\xdelall,0) circle (1.5pt);

  \fill[orange!05] (\xphy,.05) rectangle (\xdelall,3.45);
  \node[anchor=north, black] at ({(\xphy+\xdelall)/2},3.75) {\footnotesize the semantic gap};

  \draw[dashed, gray!65] (\xins,0) -- (\xins,3.45);
  \node[anchor=south, font=\scriptsize] at (\xins-.25,-0.35) {\textsc{Insert}};

  \draw[dashed, gray!65] (\xlog,0) -- (\xlog,3.45);
  \draw[dashed, gray!65] (\xphy,0) -- (\xphy,3.45);
  \draw[dashed, gray!65] (\xp2e2,0) -- (\xp2e2,3.45);
  \draw[dashed, gray!65] (\xdelall,0) -- (\xdelall,3.45);

  \draw[blue!70, line width = 1.0pt]
    (0.15,0.75)
    .. controls (0.45,0.78) and (0.60,1.05) .. (\xins,\yins);
    
 \fill[fill=black!75,draw=black]
  (\xins,\yins+0.14)
  --
  (\xins-0.16,\yins-0.09)
  --
  (\xins+0.16,\yins-0.09)
  -- cycle;
 
  \fill[black] (\xlog,\ylog) circle (1.5pt);
  
\filldraw[fill=black!75,draw=black]
  (\xp2e2,\yins+0.14)
  --
  (\xp2e2-0.16,\yins-0.09)
  --
  (\xp2e2+0.16,\yins-0.09)
  -- cycle;
  
    \fill[black] (\xdelall,0.75) circle (1.5pt);
    
  \node[font=\scriptsize, gray!60!black, anchor=west] (prelab) at (1,\yins+0.85)
    {\tiny{pre-insertion reference point}};
    
  \draw[-{Latex}, line width=0.5pt, gray!60] (prelab.west) -- (0.98,\yins+0.18);

  \draw[line width=1.0pt,blue!70] (\xins,\yins) -- (\xins,3.25);
  
  \draw[line width=1.0pt,blue!70] (\xins,3.25) -- (\xlog,3.25);
  
  \node[gray!70!black, font=\scriptsize] at ({(\xins+\xlog)/2},3.43) {direct access};

\draw[line width=1.0pt, blue!70]
  (\xlog,\ylog) -- (\xphy,\ylog)
  coordinate (Pphy)
  -- (\xphy,\yphy);
\node[anchor=west] at (2.35,3) {\scriptsize logical-only};
\fill[black] (\xphy,\yphy) circle (1.5pt);
   
  \draw[line width=0.8pt, dashed, gray!65] (\xins,\yins) -- (\xp2e2,\yins);

  \draw[line width=1pt, dashed, blue!70]
    (\xphy, \yphy) -- (\xsem,1.70) -- (\xp2e2,\yins);

    \node[anchor=west, font=\tiny, blue!75!black, align=left] at (5.5,2.0){gradually decreasing\\re-inference};

  \draw[line width=1.0pt, dashed, blue!70]
    (\xp2e2,\yins) -- (\xdelall,0.75);

  \node[anchor=west] at (4.10,1.05) {};

  \node[anchor=north, font=\tiny, align=left] at (\xlog,-0.33) {%
    \texttt{DELETE}\\
    \texttt{CASCADE}~\cite{postgres-fk-doc}\\
    MVCC~\cite{postgres-mvcc-intro-doc}\\
    tombstones~\cite{Lethe}
  };
  \node[font=\scriptsize, anchor=south] at (\xlog-.25,-0.35) {\textsc{Logical Delete}};

  \node[anchor=north, font=\tiny, align=left] at (\xphy,-0.33) {%
    VACUUM~\cite{postgres-sql-vacuum-doc}\\
    compaction~\cite{Lethe}\\
    deadlines~\cite{Sarkar2022}
  };
  \node[font=\scriptsize, anchor=south] at (\xphy+.25,-0.35) {\textsc{Physical Delete}};

  \node[font=\tiny, black, anchor=north] at (\xp2e2,-0.33)
    {P2E2~\cite{Chakraborty2024}};

\node[font=\scriptsize, anchor=south, align=center]
  at (\xdelall-.20,-0.60)
  {\textsc{Delete All}\\ \textsc{}};
  \node[anchor=north, font=\tiny, align=left] at (\xdelall,-0.33) {%
    DELF~\cite{DelFB}\\
    K9db~\cite{DakAlbab2023}
  };

  \path (0,-1.05);


\node[font=\scriptsize, rotate=90, red!75!black, anchor=west] at (\xdelall+.75,\yins+0.5)
  {\tiny{\#~Additional deletions}};

\def\ycLow{2.45}   
\def\ycP{3.20}     
\def\ycHigh{3.5}  

\draw[line width=1pt, red!75!black, dashed]
  (\xphy,\ycLow)
  .. controls (\xsem,2.55) and (\xp2e2-0.15,2.62) .. (\xp2e2,\ycP)
  coordinate (Ptop);

\draw[dashed, line width=1pt, red!75!black]
  (\xp2e2,\ycP) -- (\xdelall,\ycHigh);

\fill[red!75!black] (\xp2e2,\ycP) circle (1.5pt);
\fill[red!75!black] (\xdelall,\ycHigh) circle (1.5pt);

\node[
  anchor=south,
  font=\tiny,
  text=red!75!black,
  align=center
] at (\xdelall,\ycHigh) {delete all\\user related data};

\node[font=\tiny, red!75!black, anchor=south] at ({(\xp2e2+\xdelall)/2},2.82)
  {};

\node[anchor=west, font=\tiny, red!75!black, align=left] at (4.25,2.8)
  {gradually increasing\\\#~deletions};

  \end{scope}
\end{tikzpicture}
\caption{Inference timeline \&~\#additional deletions .}
\label{fig:timeline}
\end{figure}
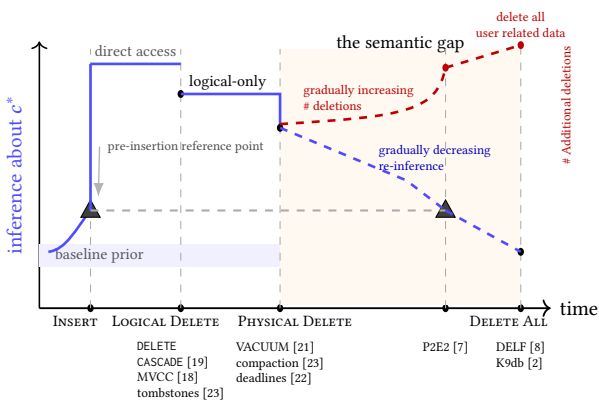

\section{Mind the Semantic Gap}
\label{sec:timeline}

Fig.~\ref{fig:timeline} makes explicit the stages implicit in our discussion on Ex.~\ref{ex:employee}. It presents deletion as an \emph{inference timeline}, from insertion through logical and physical deletion to semantic
mitigation. The central question is not only whether the deleted value is
accessible, but how much inferential support remains after logical/physical deletion.

\smallskip\noindent\textbf{Pre-insertion indirect inference.}
The timeline begins before insertion. Even if the value had never been
stored, some information about it may already be inferable from the system
state and other observations. Fig.~\ref{fig:timeline} captures this as
the \emph{pre-insertion reference point} level at \textsc{Insert}. This
gives a natural rollback target: remove the additional inferability
introduced by storage, without claiming to erase side information outside
the database.

\smallskip\noindent\textbf{Logical and physical delete.}
At \textsc{Logical Delete}, mechanisms such as \texttt{DELETE},
\texttt{CASCADE}, and MVCC visibility rules make $c^*$ disappear from query
results~\cite{postgres-fk-doc,postgres-mvcc-intro-doc}. This prevents direct
retrieval, but not inference: in Ex.~\ref{ex:employee}, dependencies D1--D3
still allow \textsf{SalaryBand} to be inferred from the remaining attributes,
and derived artifacts may preserve comparable signals. At \textsc{Physical Delete},
the system removes the bytes corresponding to $c^*$. In MVCC engines, this may
occur only when reclamation runs, e.g., \texttt{VACUUM} in
PostgreSQL~\cite{postgres-mvcc-intro-doc,postgres-sql-vacuum-doc}; in LSM-based
engines, deletion becomes durable only after compaction~\cite{Lethe}. Recent
systems expose this delay as part of the deletion guarantee~\cite{BuildingDelCompDB,Sarkar2022}.

\smallskip\noindent\textbf{The semantic gap.}
Physical deletion alone does not guarantee meaningful erasure. Once the bytes of
$c^*$ are removed, the value itself may still be recoverable from remaining
dependencies, derived artifacts, or deletion traces. We refer to this gap where storage-level deletion has
completed, but inference-level deletion has not, as the  \emph{semantic gap} (shaded orange in Fig.~\ref{fig:timeline}).

\smallskip\noindent\textbf{Rollback targets and cost.}
The blue dashed line in Fig.~\ref{fig:timeline} represents
\emph{inference-controlled deletion}: after physical deletion, the system
further actively reduces support for inferring $c^*$. The marker labelled
\textsf{P2E2} denotes insertion-referenced
rollback~\cite{Chakraborty2024}: reduce post-deletion inference to the
pre-insertion indirect inference level, removing the incremental
inferability introduced by storing $c^*$ without claiming to erase what was
already inferable before insertion. At the other extreme, \textsf{DELF} and
\textsf{K9db} illustrate broad deletion across user-related data and
stores~\cite{DelFB,DakAlbab2023}.
The red dashed line in Fig.~\ref{fig:timeline} shows this trade-off schematically.
This mechanism can reduce evidence about $c^*$
further.
The space between these endpoints is, in our view, the most
promising: \emph{bounded-leakage} mechanisms that trade a controlled
increase in inferability for a substantial reduction in auxiliary deletion
cost.

\smallskip\noindent\textbf{Observation dimensions.}
The timeline also brings forth the importance of an adversary’s observation dimension.
Fig.~\ref{fig:timeline} suggests three surfaces: a
\emph{forensic/storage dimension}, spanning the interval between logical and
physical deletion and motivating timeliness guarantees for physically removing the bits associated to the data~
\cite{postgres-sql-vacuum-doc,Lethe,Sarkar2022}; a \emph{logical/query
dimension}, on which inference persists through remaining dependencies; a \emph{derived-artifact/trace dimension}, on which outputs,
refresh events, or traces may still leak predictive evidence. A deletion
contract must therefore specify both when bytes disappear and what
inferential support may remain.
In the next section, we take a closer look at the semantic gap.

\section{Analyzing the Semantic Gap}
\label{sec:surfaces}


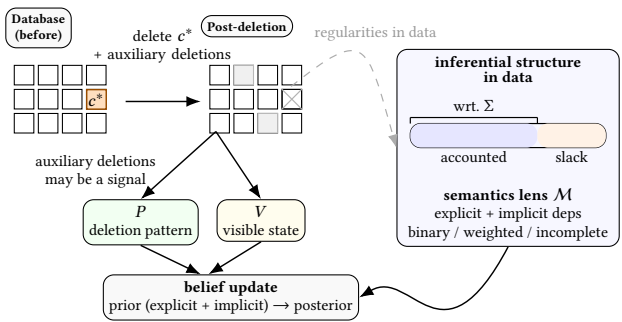
\begin{figure}[b]
\centering
\begin{tikzpicture}[x=1mm,y=1mm, font=\scriptsize]
\tikzset{
  box/.style={draw, rounded corners, inner sep=2.2pt, align=center},
  arr/.style={-Latex, line width=0.55pt},
  darr/.style={-Latex, dashed, line width=0.45pt, gray!65},
  cell/.style={draw, minimum width=2.6mm, minimum height=2.6mm, inner sep=0pt},
  tgt/.style={cell, fill=orange!25, draw=orange!60!black, line width=0.55pt},
  del/.style={cell, fill=gray!12, draw=gray!55},
}

\node[box, anchor=north west, fill=gray!5, align=center, font=\tiny] (db0) at (0,.5)
  {\textbf{Database} \\ \textbf{(before)}};
\node[anchor=north west] (tab0) at (0,-6) {%
\begin{tikzpicture}[font=\scriptsize]
  \foreach \r in {0,1,2} {
    \foreach \c in {0,1,2,3} {
      \node[cell] at (\c*3.2,-\r*3.2) {};
    }
  }
  \node[tgt] at (3.2*3,-3.2*1) {$c^*$};
\end{tikzpicture}%
};

\node[box, anchor=north west, fill=gray!5,font=\tiny] (db1) at (26,-0.5) {\textbf{Post-deletion}};
\node[anchor=north west] (tab1) at (26,-6) {%
\begin{tikzpicture}[font=\scriptsize]
  \foreach \r in {0,1,2} {
    \foreach \c in {0,1,2,3} {
      \node[cell] at (\c*3.2,-\r*3.2) {};
    }
  }
  \node[cell] (rm) at (3.2*3,-3.2*1) {};
  \draw[line width=0.45pt, gray!60] (rm.south west) -- (rm.north east);
  \draw[line width=0.45pt, gray!60] (rm.north west) -- (rm.south east);

  \node[del] at (3.2*1,-3.2*0) {};
  \node[del] at (3.2*2,-3.2*2) {};
\end{tikzpicture}%
};

\draw[arr] (16,-12) -- (26,-12);
\node[anchor=south, align=center, text width=18mm] at (21,-8) {delete $c^*$\\+ auxiliary deletions};

\node[box, anchor=north west, fill=blue!3] (semcol) at (52,-5) {%
\begin{minipage}{28mm}
\centering
\textbf{inferential structure}\\[-0.2em]
\textbf{in data}\\[0.6em]

\begin{tikzpicture}[x=1mm,y=1mm, font=\scriptsize]
  \coordinate (I0) at (0,0);
  \coordinate (I1) at (26,0);
  \coordinate (IS) at ($(I0)!0.65!(I1)$);

  \draw[line width=0.45pt] ($(I0)+(0,-1.6)$) rectangle ($(I1)+(0,1.6)$);
  \fill[blue!10] ($(I0)+(0,-1.6)$) rectangle ($(IS)+(0,1.6)$);
  \fill[orange!10] ($(IS)+(0,-1.6)$) rectangle ($(I1)+(0,1.6)$);

  \node[anchor=north, yshift=-1.2mm] at ($(I0)!0.5!(IS)$) {accounted};
  \node[anchor=north, yshift=-1.2mm] at ($(IS)!0.5!(I1)$) {slack};

  \draw[line width=0.45pt] ($(I0)+(0,2.4)$) -- ($(IS)+(0,2.4)$);
  \draw[line width=0.45pt] ($(I0)+(0,2.4)$) -- ($(I0)+(0,1.6)$);
  \draw[line width=0.45pt] ($(IS)+(0,2.4)$) -- ($(IS)+(0,1.6)$);
  \node[anchor=south] at ($(I0)!0.5!(IS)+(0,2.4)$) {\scriptsize wrt.\ $\Sigma$};
\end{tikzpicture}

\vspace{1.8mm}
\textbf{semantics lens} $\mathcal{M}$\\
\scriptsize explicit + implicit deps\\
\scriptsize binary / weighted / incomplete
\end{minipage}%
};

\draw[darr] (38,-10) .. controls (48,-2) and (48,-4) .. (semcol.west);
\node[gray!80, anchor=west] at (40,-3) {regularities in data};

\node[box, anchor=north, fill=green!5] (V) at (18,-25) {\textbf{$P$}\\deletion pattern};
\node[box, anchor=north, fill=yellow!8] (P) at (34,-25) {\textbf{$V$}\\visible state};

\draw[arr] (28,-16) -- (V.north);
\draw[arr] (28,-16) -- (P.north);

\node[anchor=north, align=center] at (12,-18) {auxiliary deletions\\may be a signal};

\node[box, anchor=north, fill=gray!5] (belief) at (30,-35)
{\textbf{belief update}\\prior (explicit + implicit) $\rightarrow$ posterior};

\draw[arr] (V.south) -- (26,-35);
\draw[arr] (P.south) -- (27,-35);

\draw[arr] (semcol.south) .. controls (58,-48) and (54,-34) .. (belief.east);

\end{tikzpicture}
\caption{The Semantic Gap.}
\label{fig:vp-pipeline}
\end{figure}

We now analyze what an adversary observes after logical/physical deletion, how those
observations support inference, and how guarantees should be stated relative
to an explicit semantics model.

\smallskip\noindent\textbf{Residual state and deletion pattern.}
Consider a deletion request for a sensitive cell $c^*$. A mechanism
transforms the database and may choose an auxiliary deletion set $S$ besides $c^*.$
After the mechanism executes, the adversary sees two observables.
The first is the \emph{residual state} $V$: the data that remains visible
after deletion, including base tables and any externally visible derived
artifacts such as aggregates. The second is the \emph{deletion pattern} $P$: what the deletion
operation itself reveals. In the simplest case, $P$ includes the auxiliary
set $S$. More generally, it may include coarser summaries, such as the size
of $S$ or its attribute footprint. Figure~\ref{fig:vp-pipeline} depicts the two leakage channels.

\emph{Residual-state leakage} occurs when the remaining state $V$ still
supports inference about $c^*$. Example~\ref{ex:employee} shows this
directly: even after removing the target value, the remaining attribute values may
still reveal it through dependencies.

\emph{Pattern leakage} occurs when the mitigation pattern itself is
informative. In dependency-rich settings, different inference paths may lead
to different auxiliary deletion sets. An adversary who observes $P$ can
therefore infer which path was active, even if the residual state $V$
appears plausible. Related ideas appear in work on \emph{deceptive
deletions}, where deletion events are treated as signals and decoys are used
to reduce adversarial advantage~\cite{minaei-ndss21-deceptive}. For our
setting, the relevant observation is that reducing residual-state leakage may make the
deletion pattern more revealing.

Inference is not always restricted to tuples. Aggregates, summaries, and released
outputs can also support inference, even when direct access is absent, a
classic concern in statistical database security and inference control
\cite{needham-inference-control,adam89-statdb}. Here, such channels
belong to $V$ when they appear in released outputs, and may also affect $P$
when deletions trigger visible refreshes or changes in aggregate behavior.

\smallskip\noindent\textbf{Bounded belief updates.}
A privacy-preserving deletion guarantee need not require the adversary's posterior
belief about $c^*$ to be small, which may often be unrealistic. The
adversary may already have a strong prior based on external information.
What matters is how much the system changes that belief.

Let $\pi$ be the adversary's prior over $c^*$, and let $\pi'$ be the
posterior after observing $(V,P)$. A bounded-leakage contract constrains the
update from $\pi$ to $\pi'$. This offers a natural way to formalize
semantic rollback in Figure~\ref{fig:timeline}: the goal is to limit the
additional evidence the system provides after deletion, relative to a stated
reference point such as pre-insertion inference~\cite{Chakraborty2024}.

This view is also close in spirit to differential privacy~(DP), which controls
how much an observation can change an adversary's beliefs, regardless of
prior side information~\cite{Dwork2014}. DP provides a useful reference point for
bounded-leakage deletion: the contract should bound posterior shifts induced
by $(V,P)$, regardless of the deleted value, while minimizing the cost of
auxiliary deletions.

\smallskip\noindent\textbf{Where semantics fit.}
The observables $(V,P)$ matter only through the semantics an adversary uses
to interpret them. We model this through a semantics model $\mathcal{M}$: a
set of dependencies or inference rules that connect observable state to the
deleted cell.

This immediately raises a coverage question: does $\mathcal{M}$ account for
all inference routes that matter, or only the subset represented by the
system? The accounted/slack split suggests two complementary formalizations
of deletion guarantees.

\emph{Accounted guarantees (model-relative).}
One can state guarantees \emph{relative to} the modeled semantics lens
$\mathcal{M}$ (or its represented dependency set $\Sigma$). Here, the
adversary is restricted to inference routes expressible in $\Sigma$, and the
guarantee bounds belief shifts induced by $V$ and $P$ under that model.
Such a claim is meaningful when $\Sigma$ is the object the system can
enforce, and it makes explicit that the guarantee is conditional on the
model being a sound approximation of the relevant inference paths.

\emph{Slack-aware guarantees (model-mismatch).}
Alternatively, one can treat slack as an explicit model-mismatch term. Let
$\Sigma^\star$ denote an idealized dependency set capturing the true
inferential structure in the data, and write
$\Sigma\subseteq\Sigma^\star$ for the modeled subset. A slack-aware
guarantee then decomposes the total belief shift into a controlled component
(attributable to $\Sigma$) and a residual component (attributable to
$\Sigma^\star\!\setminus\!\Sigma$). This motivates a parameterized guarantee
of the form: $\text{Leak}(V,P;\Sigma^\star)
\;\le\;
\text{Leak}(V,P;\Sigma)
\;+\;
\text{Slack}(V,P;\Sigma^\star\!\setminus\!\Sigma),$
where the first term is enforced by the mechanism, and the second is either
bounded (e.g., via coverage assumptions or sampling confidence on mined
rules) or reported as residual risk. 

Figure~\ref{fig:taxonomy-bands} presents the design space of deletion (and 
mechanisms);
\textbf{the side panel summarizes how to read the figure}. The open areas (Areas~1,2,\& 3) motivate the open challenges.


\definecolor{cT1}{RGB}{214,232,248}
\definecolor{cT2}{RGB}{232,218,244}
\definecolor{cT3}{RGB}{210,236,210}
\definecolor{cT4}{RGB}{255,243,210}
\definecolor{cT5}{RGB}{248,212,212}
\definecolor{cT6}{RGB}{210,236,230}
\definecolor{cI1}{RGB}{235,225,215}
\definecolor{cI2}{RGB}{220,220,235}
\definecolor{cGap}{RGB}{253,212,158}
\definecolor{cDiv}{RGB}{160,160,160}
\definecolor{cBdr}{RGB}{70,70,70}
\definecolor{cSec}{RGB}{50,50,50}
\definecolor{cPanelBg}{RGB}{248,248,248}
\definecolor{cGroupA}{RGB}{242,242,242}
\definecolor{cGroupB}{RGB}{232,232,232}

\tikzset{
  axlbl/.style={font=\small\bfseries, text=cBdr,
                align=center, text width=1.90cm},
  bktlbl/.style={font=\small\bfseries, text=black, align=center},
  bktlblfade/.style={font=\small\bfseries, text=black, align=center},
  syslbl/.style={font=\scriptsize, text=black!75, align=center, inner sep=0pt, line width=0pt},
  gaplbl/.style={font=\small\itshape, text=cBdr!80, align=center},
  notelbl/.style={font=\scriptsize\itshape, text=black,
                  align=center, text width=10.5cm},
  grouplbl/.style={font=\small\bfseries, text=cBdr, align=center},
  openbadge/.style={
    draw=orange!65, fill=orange!10, rounded corners=1.5pt,
    line width=0.4pt, font=\scriptsize\bfseries,
    inner sep=1.5pt, text=orange!70!black, align=center},
  paneltitle/.style={font=\small\bfseries, text=black, align=left},
  panelhead/.style={font=\small\bfseries, text=black!85, align=left},
  paneltxt/.style={font=\small, text=black!78, align=left},
  panelbox/.style={draw=cBdr!35, fill=cPanelBg, rounded corners=2pt, line width=0.45pt}
}

\begin{figure*}[t]
\centering
\begin{tikzpicture}[x=1.00cm, y=0.92cm]

\path[use as bounding box] (-0.03,-0.02) rectangle (18.00,11.20);

\def\xGroup{0.95}
\def\xLabel{3.30}
\def\xAxis{2.125}

\def\xA{8.55}
\def\xB{13.50}
\def\xMidTwo{10.90}

\def\xCOne{5.925}
\def\xCTwo{11.025}
\def\xCThree{15.75}
\def\xDOne{7.10}
\def\xDTwo{14.45}

\def\Wtotal{18.00}

\def\rIIbot{0.00}
\def\rIItop{1.00}
\def\rIbot{1.00}
\def\rItop{2.00}

\def\rSplit{2.20}

\def\rTsixBot{2.20}
\def\rTsixTop{3.70}
\def\rTfiveBot{3.70}
\def\rTfiveTop{5.20}
\def\rTfourBot{5.20}
\def\rTfourTop{6.70}
\def\rTthreeBot{6.70}
\def\rTthreeTop{8.20}
\def\rTtwoBot{8.20}
\def\rTtwoTop{9.70}
\def\rToneBot{9.70}
\def\rToneTop{11.20}

\pgfmathsetmacro{\mTone}{(\rToneBot+\rToneTop)/2}
\pgfmathsetmacro{\mTtwo}{(\rTtwoBot+\rTtwoTop)/2}
\pgfmathsetmacro{\mTthree}{(\rTthreeBot+\rTthreeTop)/2}
\pgfmathsetmacro{\mTfour}{(\rTfourBot+\rTfourTop)/2}
\pgfmathsetmacro{\mTfive}{(\rTfiveBot+\rTfiveTop)/2}
\pgfmathsetmacro{\mTsix}{(\rTsixBot+\rTsixTop)/2}
\pgfmathsetmacro{\mI}{(\rIbot+\rItop)/2}
\pgfmathsetmacro{\mII}{(\rIIbot+\rIItop)/2}

\fill[cT1]  (0,\rToneBot) rectangle (\Wtotal,\rToneTop);
\fill[cT2]  (0,\rTtwoBot) rectangle (\Wtotal,\rTtwoTop);
\fill[cGap] (\xA,\rTtwoBot) rectangle (\Wtotal,\rTtwoTop);

\fill[cT3]  (0,\rTthreeBot) rectangle (\Wtotal,\rTthreeTop);
\fill[cGap] (\xB,\rTthreeBot) rectangle (\Wtotal,\rTthreeTop);

\fill[cT4] (0,\rTfourBot) rectangle (\Wtotal,\rTfourTop);
\fill[cT5] (0,\rTfiveBot) rectangle (\Wtotal,\rTfiveTop);
\fill[cT6] (0,\rTsixBot) rectangle (\Wtotal,\rTsixTop);

\fill[cI1] (0,\rIbot) rectangle (\Wtotal,\rItop);
\fill[cI2] (0,\rIIbot) rectangle (\Wtotal,\rIItop);

\fill[cGroupA] (0,\rSplit) rectangle (\xGroup,\rToneTop);
\fill[cGroupB] (0,\rIIbot) rectangle (\xGroup,\rSplit);

\pgfmathsetmacro{\grpIntrMid}{(\rSplit+\rToneTop)/2}
\pgfmathsetmacro{\grpInvMid}{(\rIIbot+\rSplit)/2}
\node[grouplbl, rotate=90] at (0.475,\grpIntrMid) {Intrinsic properties};
\node[grouplbl, rotate=90, align=center] at (0.475,\grpInvMid) {Invocation\\context};

\draw[cDiv,line width=0.45pt](\xGroup,\rIIbot)--(\xGroup,\rToneTop);
\draw[cDiv,line width=0.45pt](\xLabel,\rIIbot)--(\xLabel,\rToneTop);

\draw[cDiv,line width=0.35pt](\xA,\rToneBot)--(\xA,\rToneTop);
\draw[cDiv,line width=0.35pt](\xB,\rToneBot)--(\xB,\rToneTop);
\draw[cDiv,line width=0.35pt](\xA,\rTtwoBot)--(\xA,\rTtwoTop);
\draw[cDiv,line width=0.35pt](\xB,\rTtwoBot)--(\xB,\rTtwoTop);
\draw[cDiv,line width=0.35pt](\xA,\rTthreeBot)--(\xA,\rTthreeTop);
\draw[cDiv,line width=0.35pt](\xB,\rTthreeBot)--(\xB,\rTthreeTop);
\draw[cDiv,line width=0.35pt](\xMidTwo,\rTfourBot)--(\xMidTwo,\rTfourTop);
\draw[cDiv,line width=0.35pt](\xMidTwo,\rTfiveBot)--(\xMidTwo,\rTfiveTop);
\draw[cDiv,line width=0.35pt](\xA,\rTsixBot)--(\xA,\rTsixTop);
\draw[cDiv,line width=0.35pt](\xB,\rTsixBot)--(\xB,\rTsixTop);
\draw[cDiv,line width=0.35pt](\xMidTwo,\rIbot)--(\xMidTwo,\rItop);
\draw[cDiv,line width=0.35pt](\xMidTwo,\rIIbot)--(\xMidTwo,\rIItop);

\node[axlbl] at(\xAxis,\mTone)   {C1\\Rollback\\objective};
\node[axlbl] at(\xAxis,\mTtwo)   {C2\\Observables\\controlled};
\node[axlbl] at(\xAxis,\mTthree) {C3\\Semantics\\model $\mathcal{M}$};
\node[axlbl] at(\xAxis,\mTfour)  {C4\\Recoverability};
\node[axlbl] at(\xAxis,\mTfive)  {C5\\Expiry-\\awareness};
\node[axlbl] at(\xAxis,\mTsix)   {C6\\Implementation\\layer};
\node[axlbl] at(\xAxis,\mI)      {I1\\Initiator};
\node[axlbl] at(\xAxis,\mII)     {I2\\Trigger};

\pgfmathsetmacro{\hdrTone}{\rToneTop-0.35}
\pgfmathsetmacro{\sysYone}{(\rToneBot+\mTone)/2+0.02}

\node[bktlbl] at(\xCOne,\hdrTone)   {\textsc{Logical}};
\node[bktlbl] at(\xCTwo,\hdrTone)   {\textsc{Physical}};
\node[bktlbl] at(\xCThree,\hdrTone) {\textsc{Semantic}};

\node[syslbl,text width=4.60cm] at(\xCOne,\sysYone){%
  \textsc{delete}/\textsc{cascade}~\cite{postgres-fk-doc},
  Mongo TTL~\cite{mongodb-ttl}, DELF~\cite{DelFB},\\
  K9db~\cite{DakAlbab2023}, Edna~\cite{edna-sosp23}};

\node[syslbl,text width=4.50cm] at(\xCTwo,\sysYone){%
  Lethe~\cite{Lethe}, Timely del.~\cite{Sarkar2022},
  Cassandra~\cite{Cassandra}};

\node[syslbl,text width=4.30cm] at(\xCThree,\sysYone){%
  P2E2~\cite{Chakraborty2024}};

\pgfmathsetmacro{\hdrTtwo}{\rTtwoTop-0.35}
\pgfmathsetmacro{\sysYtwoMid}{\mTtwo-0.12}

\node[bktlbl] at(\xCOne,\hdrTtwo)   {\textsc{Remaining visible data ($V$)}};
\node[bktlbl] at(\xCTwo,\hdrTtwo)   {\textsc{Deletion pattern ($P$)}};
\node[bktlbl] at(\xCThree,\hdrTtwo) {\textsc{Both ($V{+}P$)}};

\node[syslbl,text width=4.60cm] at(\xCOne,\sysYtwoMid){%
  \textsc{delete}/\textsc{cascade}~\cite{postgres-fk-doc},
  Mongo TTL~\cite{mongodb-ttl},
  DELF~\cite{DelFB},\\
  K9db~\cite{DakAlbab2023},
  Edna~\cite{edna-sosp23},
  Lethe~\cite{Lethe}, Timely del.~\cite{Sarkar2022},
  Cassandra~\cite{Cassandra}, P2E2~\cite{Chakraborty2024}};

\node[gaplbl]    at(\xCTwo,\mTtwo)    {Area~1};
\node[openbadge] at(\xCTwo,\sysYtwoMid-.25) {open};
\node[gaplbl]    at(\xCThree,\mTtwo)    {Area~2};
\node[openbadge] at(\xCThree,\sysYtwoMid-.25) {open};

\pgfmathsetmacro{\hdrTthree}{\rTthreeTop-0.35}
\pgfmathsetmacro{\sysYthree}{\mTthree-0.20}

\node[bktlbl] at(\xCOne,\hdrTthree)   {\textsc{Slack}};
\node[bktlbl] at(\xCTwo,\hdrTthree)   {\textsc{Binary (0/1)}};
\node[bktlbl] at(\xCThree,\hdrTthree) {\textsc{Weighted}};

\node[syslbl,text width=4.60cm] at(\xCOne,\sysYthree){%
  Mongo TTL~\cite{mongodb-ttl}, Lethe~\cite{Lethe},\\
  Timely del.~\cite{Sarkar2022},
  Cassandra~\cite{Cassandra}};

\node[syslbl,text width=4.50cm] at(\xCTwo,\sysYthree){%
  \textsc{delete}/\textsc{cascade}~\cite{postgres-fk-doc},
  DELF~\cite{DelFB}, K9db~\cite{DakAlbab2023},\\
  Edna~\cite{edna-sosp23}, P2E2~\cite{Chakraborty2024}};

\node[gaplbl]    at(\xCThree,\mTthree) {Area 3};
\node[openbadge] at(\xCThree,\sysYthree-.25) {open};

\pgfmathsetmacro{\hdrTfour}{\rTfourTop-0.35}
\pgfmathsetmacro{\sysYfour}{\mTfour-0.18}

\node[bktlbl] at(\xDOne,\hdrTfour){\textsc{Irreversible}};
\node[bktlbl] at(\xDTwo,\hdrTfour){\textsc{Reversible window}};

\node[syslbl,text width=7.60cm] at(\xDOne,\sysYfour){%
  \textsc{delete}/\textsc{cascade}~\cite{postgres-fk-doc},
  Mongo TTL~\cite{mongodb-ttl}, K9db~\cite{DakAlbab2023},
  Lethe~\cite{Lethe}, Timely del.~\cite{Sarkar2022},
  Cassandra~\cite{Cassandra}, P2E2~\cite{Chakraborty2024}};

\node[syslbl,text width=7.00cm] at(\xDTwo,\sysYfour){%
  DELF~\cite{DelFB} \textit{(temp.\ reversible)},
  Edna~\cite{edna-sosp23} \textit{(reveal)}};

\pgfmathsetmacro{\hdrTfive}{\rTfiveTop-0.35}
\pgfmathsetmacro{\sysYfive}{\mTfive-0.18}

\node[bktlbl] at(\xDOne,\hdrTfive){\textsc{External (caller)}};
\node[bktlbl] at(\xDTwo,\hdrTfive){\textsc{Native (engine)}};

\node[syslbl,text width=7.60cm] at(\xDOne,\sysYfive){%
  \textsc{delete}/\textsc{cascade}~\cite{postgres-fk-doc},
  DELF~\cite{DelFB},\\
  K9db~\cite{DakAlbab2023}, Edna~\cite{edna-sosp23}};

\node[syslbl,text width=7.00cm] at(\xDTwo,\sysYfive){%
  Mongo TTL~\cite{mongodb-ttl}, Lethe~\cite{Lethe},\\
  Timely del.~\cite{Sarkar2022}, Cassandra~\cite{Cassandra},
  P2E2~\cite{Chakraborty2024}};

\pgfmathsetmacro{\hdrTsix}{\rTsixTop-0.35}
\pgfmathsetmacro{\sysYsix}{\mTsix-0.18}

\node[bktlbl] at(\xCOne,\hdrTsix)   {\textsc{LSM-tree}};
\node[bktlbl] at(\xCTwo,\hdrTsix)   {\textsc{MVCC/heap}};
\node[bktlbl] at(\xCThree,\hdrTsix) {\textsc{App-layer}};

\node[syslbl,text width=4.60cm] at(\xCOne,\sysYsix){%
  Lethe~\cite{Lethe}, Cassandra~\cite{Cassandra}};

\node[syslbl,text width=4.50cm] at(\xCTwo,\sysYsix){%
  \textsc{delete}/\textsc{cascade}~\cite{postgres-fk-doc},
  Timely del.~\cite{Sarkar2022}};

\node[syslbl,text width=4.30cm] at(\xCThree,\sysYsix){%
  Mongo TTL~\cite{mongodb-ttl}, DELF~\cite{DelFB},\\
  K9db~\cite{DakAlbab2023}, Edna~\cite{edna-sosp23},
  P2E2~\cite{Chakraborty2024}};

\node[bktlblfade] at(\xDOne,\mI)   {\textsc{Data subject}};
\node[bktlblfade] at(\xDTwo,\mI) {\textsc{System / policy}};
\node[notelbl]    at(\xMidTwo,\mI-0.30) {%
  All mechanisms support any initiator, depending on deployment.};

\node[bktlblfade] at(\xDOne,\mII)   {\textsc{Demand-driven}};
\node[bktlblfade] at(\xDTwo,\mII) {\textsc{Retention-driven}};
\node[notelbl]    at(\xMidTwo,\mII-0.30) {%
  Any mechanism can be scheduled; T5 captures native expiry support.};

\draw[cBdr,line width=0.7pt](0,\rIIbot)rectangle(\Wtotal,\rToneTop);

\foreach \y in {\rTsixTop,\rTfiveTop,\rTfourTop,\rTthreeTop,\rTtwoTop}{
  \draw[white,   line width=1.3pt] (\xGroup,\y)--(\Wtotal,\y);
  \draw[cBdr!22, line width=0.35pt](\xGroup,\y)--(\Wtotal,\y);
}
\draw[white,   line width=1.3pt] (\xGroup,\rIItop)--(\Wtotal,\rIItop);
\draw[cBdr!22, line width=0.35pt](\xGroup,\rIItop)--(\Wtotal,\rIItop);

\draw[white,   line width=1.3pt] (0,\rSplit)--(\Wtotal,\rSplit);
\draw[cBdr!28, line width=0.45pt](0,\rSplit)--(\Wtotal,\rSplit);

\end{tikzpicture}

\vspace{0.6em}
\begin{tikzpicture}
\node[
  draw=cBdr!35, fill=cPanelBg, rounded corners=3pt,
  line width=0.45pt,
  text width=\linewidth-16pt,  
  font=\small,
  align=left,
  inner xsep=8pt,
  inner ysep=8pt
] {%
  \textbf{\textsc{Reading the taxonomy}}\newline
  \textbf{Structure:}\enspace
  C1--C6 capture \emph{intrinsic} properties of a deletion mechanism;
  I1--I2 are \emph{invocation-context} overlays.\newline
  \textbf{Interpretation:}\enspace
  The upper axes describe \emph{what} a mechanism rolls back, controls,
  and assumes semantically; the lower axes describe \emph{how} it operates
  in practice.\newline
  \textbf{Open areas:}\enspace
  Most systems control only $V$, not $P$.\enspace
  Joint control of $(V,P)$ remains largely open.\enspace
  Most semantic mechanisms rely on binary models; weighted
  semantics remain open.\quad
  \newline
  \textbf{Takeaway:}\enspace
  The taxonomy is not only a classification of existing systems; it also
  marks the open regions that motivate the agenda that follows.
};
\end{tikzpicture}

\caption{Taxonomy \& Design Space.}
\label{fig:taxonomy-bands}
\end{figure*}



\section{Open Challenges}
\label{sec:open}
 Fig.~\ref{fig:taxonomy-bands} shows: most systems control only the residual state $V$, and conservatively over-delete to ensure no leakage due to $P$; most semantic mechanisms rely on binary
dependency models. 

In particular, the lack of mechanisms that control deletion observables beyond residual state $V$~(Areas~1~\&~2) motivates \textbf{R1}, which broadens attention from logical deletion patterns $P$ to physical maintenance traces. The absence of weighted semantic models~(Area~3) motivates \textbf{R2}, while \textbf{R3} recognizes that such semantic guarantees must be maintained as data and artifacts evolve. Finally, because these guarantees span multiple components and observation surfaces, \textbf{R4} provides the attestation layer needed to make them composable and auditable~(often required for compliance~\cite{Chakraborty2024datacase}).

\smallskip\noindent\textbf{(R1) Trace-private timeliness.}
\emph{Goal: satisfy physical timeliness guarantees while limiting what other channels' leakage reveals.}
Physical deletion is carried out through background processes such as vacuuming, compaction, tombstone garbage collection, index rebuilds, and cache refresh~\cite{postgres-routine-vacuuming-doc, Lethe}. These processes are often observable through logs, metrics, throughput changes, or maintenance events. Their timing may also correlate with deletion workload. As a result, an observer may learn the fact that a deletion occurred, and possibly the time when it occurred.

This creates a side channel that current systems do not model explicitly. One challenge is to define the right observability model for maintenance traces. Unlike $V$ or $P$, the trace is a time series of system events. A second challenge is to design mechanisms that make such traces less revealing. A deletion-aware scheduler may need to support deadline windows rather than exact deadlines, batch and coalesce requests, and inject cover work that makes deletion-induced activity harder to distinguish from background maintenance. The core question is how much randomness, padding, or reordering is enough under realistic observation models.

\smallskip\noindent\textbf{(R2) Deletion planning under weighted and scoped semantics.}
\emph{Goal: compute low-cost mitigation plans for derived artifacts under weighted dependencies and declared workload scope.} Deletion rarely stops at a base table~\cite{Kimelfeld2012, Makhija2025}. Views, caches, dashboards, feature stores, and aggregates may all carry inferential support for the deleted value. Reducing re-inference may therefore require updates or invalidations across multiple artifacts. Existing approaches either delete aggressively for safety or follow a fixed dependency model~\cite{Chakraborty2024}. Neither gives a principled way to minimize the mitigation footprint under a bounded-leakage target.

This raises three connected challenges. 
First, systems need \emph{weighted dependencies}. 
Binary models force a threshold decision: they either treat every dependency as certain, leading to excessive deletion, or ignore weak dependencies, even though many weak channels may combine to cause non-trivial leakage. 
Second, systems need \emph{workload-scoped contracts}. In practice, guarantees are often promised only with respect to a declared product surface, such as a set of views, query templates, or APIs. These scopes must be explicit and maintained as the workload evolves. 
Third, systems need \emph{planning and certification}. We need an erasure planner that chooses a low-cost set of artifact updates and invalidations subject to a leakage bound, and produces a certificate explaining why the plan satisfies that bound.

\smallskip\noindent\textbf{(R3) Maintained erasure under distribution drift.}
\emph{Goal: preserve semantic guarantees as data, dependencies, and downstream models evolve.}
Semantic deletion cannot be treated as a one-time action. A guarantee that holds at deletion time may fail later. New tuples may strengthen a previously weak dependency~\cite{DisDynamic, bleifuss2024discovering}. A model update may make the deleted value easier to predict from surviving attributes. A newly materialized artifact may introduce an inference path that did not exist when the original mitigation was computed.

Erasure can therefore be treated as a maintained property. The main difficulty is that inference risk is not monotone in data updates. New data may increase or decrease risk. Standard incremental view maintenance does not solve this problem as it tracks correctness, not leakage. A practical solution would need incremental summaries of inference strength, drift triggers that detect when a prior guarantee is no longer valid, and mitigation procedures that restore the guarantee without recomputing everything from scratch.

\smallskip\noindent\textbf{(R4) Composable deletion attestation.}
\emph{Goal: produce machine-checkable certificates that compose across database components and downstream artifacts.}
Deletion compliance is difficult to verify end-to-end. A DBMS may record that a \texttt{DELETE} was executed. A storage engine may record that bytes were reclaimed. A pipeline may record that a cache was refreshed. But these events do not automatically compose into a single claim that a target value is no longer inferable from the overall system.

This motivates a new abstraction: \emph{deletion attestation}. An attestation should name the relevant system components and state two kinds of claims: a \emph{timeliness clause}, which says when physical reclamation is completed, and a \emph{semantic clause}, which states a leakage bound under a named dependency model and observation surface. Each component should expose a local attestation that can be verified from its own metadata. The broader challenge is to make these local attestations compose without global recomputation. This requires a common semantic vocabulary and explicit interfaces between the DBMS and downstream data products.


Together, R1--R4 outline a path towards deletion as an explicit, inference-aware systems capability with well-formed, user-facing tunable privacy guarantees rather than a best-effort operation.

\section*{Acknowledgments}
Chakraborty was supported by a fellowship from HPI@UCI. 
Pandey was supported by funding from the United States–India Educational Foundation (USIEF) and the U.S. Department of State’s Fulbright Visiting Scholar Program.   
This work was supported by NSF Grants No. 2032525, 1545071, 1527536, 1952247, 2008993, 2133391, 2245372, and 2245373.
\balance
\bibliographystyle{ACM-Reference-Format}
\bibliography{sample-base}

\end{document}